\documentclass[]{mn2e}
\usepackage{psfig,subfigure,mncite}
\def\deg{$^{\circ}$}
\def\brakk#1{{\langle#1\rangle}}

\title[Spot Sizes on Sun-like Stars]
	{Spot Sizes on Sun-like Stars}

\author[S.K. Solanki \& Y.C. Unruh]
       {S.~K.~Solanki,$^1$ and Y.~C.~Unruh$^2$ \\
	$^1$ Max-Planck-Institut f\"ur Aeronomie, D-37191 Katlenburg-Lindau, Germany \\
	$^2$ Astrophysics Group, Blackett Laboratory, Imperial College for Science, Technology and Medicine, London, SW7 2BW, United Kingdom
\\
              }
\date{Accepted 4th Nov 2003; Received Aug 2003; in original form May 2003}

\begin{document}

\maketitle
\begin{abstract}
The total area coverage by starspots is of interest for a variety 
of reasons, but direct techniques only provide estimates of this important
quantity. Sunspot areas exhibit a lognormal size distribution irrespective 
of the phase of the activity cycle, implying that most sunspots 
are small. Here we explore the consequences if starspot areas 
were similarly distributed. The solar data allow 
for an increase in the fraction of larger sunspots with increasing activity.
Taking this difference between the size distribution at sunspot maximum
and minimum, we extrapolate to 
higher activity levels, assuming different
dependencies of the parameters of the lognormal distribution on total
spot coverage. We find that even for very heavily spotted (hypothetical)
stars a large fraction of the spots are smaller than the current 
resolution limit of Doppler images and might hence be missed on traditional 
Doppler maps.
\end{abstract}

\begin{keywords}
sunspots, stars: spots, stars: late-type, stars: activity
\end{keywords}
\section{Introduction} 
%
\label{sec:intro}
In recent years, an ever increasing number of spotted stars have been mapped
using Doppler imaging. The maps reveal the surface distribution of starspots, 
which in general are large compared to even the largest sunspots. While 
Doppler images do a good job of catching starspots that modulate 
the line profile, it is extremely difficult to detect a background 
of small starspots more or less homogeneously distributed over 
the stellar surface.  While TiO-band mapping still suffers from notable uncertainties
(see Sec.~2), it should in principle be able to pick up non-modulating
and homogeneous spot distributions. Typically, techniques using TiO 
bands to determine spot temperatures and surface areas tend to find 
larger covering fractions than Doppler imaging techniques
(though see also \scite{berdyugina2002potsdam}). 
Such differences in apparent spot coverage only provide a hint for
unresolved starspots. One star for which the distribution of spot 
sizes is known in great detail is the Sun. \scite{bogdan88} found the 
size distribution to be well represented by a lognormal function. This implies 
that the number of small sunspots is much larger than that of large spots. 
This supports the idea that there could be additional small, i.e. unresolved, 
starspots on more active stars as well. The total starspot coverage is 
of interest as a measure of stellar magnetic activity, in  order to establish the
proper ratio of starspot to stellar plage \cite{radick98} and to obtain 
improved estimates of the total magnetic flux carried by the star
(with the possible exception of Zeeman Doppler imaging where most of the 
magnetic signal appears to come from penumbral-type structures, techniques 
of stellar magnetic field measurement mainly sample 
plage fields \cite{saar88,solanki92cs7}).

Here we explore hypothetical scenarios for extrapolating the solar
spot-size distribution to activity levels typical for much more 
active stars. The basic assumption 
is that the size distribution of star spots can be described by a 
lognormal function, as in the case of sunspots. This assumption 
is not unreasonable since the magnetic fields on both the Sun and 
on more active cool stars are thought to be produced by a dynamo 
residing at the base of the convection zone \cite{petrovay2001,schussler2002}.
From there flux tubes carry the field to 
the solar surface. The fragmentation of these tubes during their 
passage through the convection zone is thought to give rise to the 
observed lognormal distribution \cite{bogdan88}. 
Lognormal distributions can, however, differ significantly from 
each other in their parameters. 

In order to constrain these parameters for active stars we investigate
the possible range of behaviour between solar activity minimum and maximum
and use these to extrapolate to larger levels of activity. 
Hence we assume that the processes which lead to the flux-tube 
size distribution do not change qualitatively with increasing 
activity. Such an assumption has in the past helped to reproduce, e.g., 
the high latitudes of starspots \cite{schuessler92,schussler96buoy,schrijver2001}, 
or the presence of active longitudes on the Sun and 
Sun-like stars \scite{berdyugina2003longitude}.

\begin{table}
\label{tab:starlist}
\caption[]{A selection of RS\, CVn stars whose spot coverages have been determined
using different techniques. The first column gives the name of the object, the
second and third columns the surface and spot effective temperatures according
to
\scite{oneal98tio}. The fourth column gives the inclination angle, usually taken
from the Doppler imaging papers listed in Sec.~2.1. Note that
not all groups agree on the spot temperatures and inclination angles.
}
\centerline{
   \begin{tabular}{lccc}
   \hline
   Object       & T$_{\rm star}$ & T$_{\rm spot}$       & inclination \\
   \hline
   II Peg       &       4750    & 3530                  & 60\deg        \\
   EI Eri       &       5600    & 3700                  & 46\deg        \\
   $\sigma$ Gem &       4600    & 3850                  & 60\deg        \\
   DM UMa       &       4600    & 3570                  & 55\deg        \\
   HD 199178    &       5350    & 3800                  & 40\deg        \\
   \hline
   \end{tabular}
}
\end{table}

\section{Different mapping techniques}
\label{sec:mapping}
To our knowledge, TiO modelling has so far mainly 
been published for giant stars. One exception is \scite{neff1990tio}, 
who calculate filling factors for 2 dwarf stars. As they 
do not give the epoch for their observations, comparison to Doppler
maps is difficult. We therefore limit the discussion to 5 RSCVn stars for which 
spot covering fractions have been derived from (near) simultaneous 
data using different techniques. 
Tab.~\ref{tab:starlist} lists the stars together with their 
effective and spot temperatures as well as the inclination angles 
of their rotation axes to the line of sight. 
Tab.~\ref{tab:starcover} lists the spot covering fractions for
the stars from Tab.~\ref{tab:starlist}.

The observations for the TiO {\em filling factors} have all been taken 
from \scite{neff95iipeg_tio,oneal96tio} and \scite{oneal98tio}. 
Due to activity cycles and in some cases due to incomplete phase coverage,
there are considerable variations in the filling factors measured for the 
same star by the same group, but at different times.
In order to be able to compare them to the spot coverage fractions
listed in columns 6 and 8 we converted them to a ``minimum'' and a ``most 
likely'' spot coverage. This is described in more detail later.

The measurements of the photometric spot coverages (column 6 
in Tab.~\ref{tab:starcover}) are taken from 
\scite{henry95rscvn}, \scite{rodono2000iipeg} and \scite{padmakar99} 
and are labelled (H+), R(+) and (PP), respectively, in column 7.
The covering fractions derived from photometry by \scite{henry95rscvn} are
lower limits as they have used the maximum light level during each individual
observing period to represent the brightness of the unspotted star.
If we use the brightness maxima over all of their observations, the areas
need to be increased. While the exact increase depends amongst
others on the spot geometry and the stellar and spot temperatures, we estimate
that for $\sigma$~Gem total surface coverages of about 6\% are more typical
than the values given by \scite{henry95rscvn}.
For II~Peg we find that the surface coverage was more like 15\% towards the
end of 1989 as well as during 1992 September. 
\scite{rodono2000iipeg} give two different values for the spot coverage, 
one derived using a maximum entropy method, the other (higher) one with 
a Tikhonov regularisation. When they use the theoretical maximum light-level 
inferred from TiO band calculations (see \scite{neff95iipeg_tio} 
for more details), their spot areas increased by 15\% and 20\% of the 
total surface area for the maximum entropy and Tikhonov maps respectively.

The last two columns of Table~\ref{tab:starcover} give the surface 
coverage and the references for Doppler imaging determinations. 
The covering fractions are taken from \scite{berdyugina98iipeg_di} (B98+), 
\scite{washuettl98cs10} (W98+), \scite{washuettl2001cs11} (W01+), 
\scite{dempsey1992} (D92+)\footnote{The surface coverage for 
$\sigma$~Gem derived by \scite{dempsey1992} has been obtained using 
a variant of a line-bisector analysis rather than by Doppler imaging.}, 
\scite{hatzes93siggem} (H93), \scite{hatzes95} (H95) and 
\scite{strassmeier99hd199178} (S99+). Note that most Doppler maps 
do not give the spot coverage as a direct parameter, showing
the stellar surface temperature rather than a spot filling factor. 
We were therefore only able to use a relatively small selection 
of Doppler maps where the authors had either given the spot 
coverage such as in as B98+, W98+/W01+ (priv comm) and D92+, 
or where the maps presented allowed realistic estimates. 

\begin{table*}
\label{tab:starcover}
\caption[]{Spot area coverages (for the total 
stellar surface) of the stars listed in Table~\protect{\ref{tab:starlist}}
obtained with different techniques.
The first column gives the name of the object, the second column the 
date of the observations. Columns three to five give the coverages
obtained with the TiO modelling technique. Column three gives the range
of filling factors, while columns four and five give estimates for the total
surface coverage derived from the filling factors.
Columns 6 and 7 give the total spot coverage derived from lightcurve modelling
and the corresponding reference. 
The last two columns give the surface coverage and the references for 
Doppler imaging determinations. See text for more detail.
}
\begin{tabular}{llcccclcl}
\hline 
Object  & date & \multicolumn{3}{c} {TiO bands} &
                 \multicolumn{2}{c} {Lightcurves} &
                 \multicolumn{2}{c} {Doppler imaging} \\
	&      & ff & coverage & minimum & coverage & reference & coverage & reference \\
\ \ [1] & \ \ [2] & [3]	& [4]	& [5]	& [6]	& \ \ [7] & [8]	& \ \ [9] \\
\hline 
II Peg &  Oct 89 & [43..55] &  42  &  28 &  13; 12/21 & H+; R+ &  & \\
       &  Aug 92 & [36..50] &  37  &  23 &            &        & 10-15 & B98+
\\
       &  Sep 92 & [43..56] &  43  &  28 &   9; 13/23 & H+; R+ &  & \\
       &  Jan 95 & [26..35] &  25  &  16 &  12/21     & R+     & 10-15 & B98+
\\
EI Eri &  Mar 92 & [23..36] &  25  &  11 &	      &	      &		& \\
       &  Jan 95 & [$<$12..18] & 13  &   5 &            &		& $\simeq 3$ & W98+
\\
       &  Dec 95 & [15..15] &  13  &   5 &            &		& $<7$ & W01+ \\
$\sigma$ Gem 
       &  Mar 90 & [14..26] &  19  &   9 &  2 	& H+	& 2.9-5.4 & D92+	\\
       &  Feb 91 & [27..33] &  28  &  13 &  4	& H+	&	&	\\
       &  Mar 92 & [10..20] &  14  &   6 &  3 	& H+	&  7 	& H93	\\
       &  Jan 95 & [ 3..14] &   8  &   3 & 	&	&	&	\\
       &  Dec 95 & [15..30] &  21  &  11 & 4.4 	& PP	&	&	\\
DM UMa &  Jan 95 & [30..35] & 30   & 15  & 	&	& 12	& H95 \\
HD 199178 
       & Oct 89  & [16..32] &  20  &   8 &	&	&	&	\\
       & May 89/90 & 	    &	   &	 &	&	& $\simeq 6$ & S99+ \\
\hline 
\end{tabular}
\end{table*}

Tab.~\ref{tab:starcover} indicates that Doppler imaging and photometric 
light curve modelling tend to result in a smaller spot covering fraction 
than TiO modelling\footnote{Recent work has often combined Doppler 
imaging and photometric lightcurve modelling, showing that one and 
the same starspot distribution can reproduce
both kinds of data. In some cases this has slightly increased the total spot 
coverage compared to Doppler maps alone (see, e.g., \pcite{unruh95doppler}).}.
Both, photometric light-curve modelling and 
Doppler imaging are prone to underestimating spot areas. This is 
mainly because they are not very sensitive to rotationally invariant 
surface features, e.g., banded structures or low-level and small-scale 
distributed surface features. The current
resolution limit of Doppler imaging is 3$^{\circ}$ to 5$^{\circ}$
in longitude, depending mainly on the star's rotational velocity, 
the spectrograph resolution and the signal-to-noise ratio that can be 
achieved. To put this into context, we recall that the
largest sunspots have diameters of about 1$^{\circ}$. 

In an ideal world where the spectral type of the 
unspotted parts of the target star were known with very high accuracy, 
or where the brightness and line profile of the star can be 
measured at a time when it is unspotted, the total spot area at 
other epochs can be estimated. Such prior knowledge, unfortunately, 
is in general not available. Furthermore, it is unclear whether 
rapidly rotating stars ever are free of spots. 
  
TiO modelling is the youngest technique to determine the 
spot coverage and can be used for stars of any rotation velocity
(see \scite{neff95iipeg_tio,oneal96tio} and \scite{oneal98tio} for more
detail). Either the strength or the general shape
of the TiO band heads is matched with a 
linear combination of template-star spectra at the effective 
stellar temperature and the spot temperature. In this way, the 
spot filling factor is determined. If more than one band head is observed 
it is possible to determine spot temperature and spot filling factor
independently (provided the temperature response of the band heads is 
sufficiently different from each other). The location of the spots is not 
recovered. The filling factors are weighted for the limb darkening, but
depending on the location of the spots (e.g.,~a central circular spot, or spots
close to the limb only), the area coverage can be about a factor of two 
smaller or larger than the listed filling factor.

We have therefore calculated two estimates for the area coverage of the 
stars based on the inclination of the star and the possible spans 
of surface coverages for a given filling factor as shown in figure~8 of 
\scite{oneal96tio}. The first estimate (listed in column 4 of 
Tab.~\ref{tab:starcover}) is calculated under the 
assumption that the filling factor is just the fractional 
spot coverage. For the second estimate (listed in column 5 of 
Tab.~\ref{tab:starcover}) we try and estimate a minimum coverage 
assuming that the spots are at disk centre 
where they produce the largest contribution. 

For both estimates we use the mean filling 
factor $(ff_{\rm max} + ff_{\rm min})/2$ as a starting point 
($ff_{\rm max}$ and $ff_{\rm min}$ are the largest and smallest filling 
factors observed during a given observing season). 
To obtain the minimum spot coverage 
we reduce the filling factors according to the graphs for 
disk-centre spots shown in figure~8 of \scite{oneal96tio}.  
This reduction is typically of the order of 50\%, but depends on the 
value of the filling factor. 

The average and minimum covering fractions that are obtained in this way are in fact not the
covering fractions with respect to the total stellar surface, as the
polar region that is pointing away from the observer is never visible.
As the covering fractions calculated for Doppler imaging and also
lightcurve modelling assume that the invisible part of the star is
devoid of structure, we multiply the mean and minimum coverages 
with a factor of $(1 + \sin i)/2$ and so recover the values 
listed in columns 4 and 5 of Tab.~\ref{tab:starcover}. Apart from 
the conversion from filling factor to surface spot covering, further
errors in the TiO-band modelling can be introduced because of mismatches
between the template star atmospheres and the actual stellar atmospheres. 
Note that TiO modelling assumes that stellar spectra are not 
affected by magnetic activity beyond the strengths of the 
molecular features. This is clearly a simplification and adds
uncertainty to the spot coverage fractions deduced by this 
technique.

\section{Spot distributions on the Sun}
\label{sec:dist_sun}
\scite{bogdan88} measured the size distribution of the sunspot umbral areas
recorded at Mt Wilson between 1921 and 1982. They found that the size distribution 
could be well fitted with a lognormal distribution of the form 
\begin{equation}
\label{eq:dist}
\frac{dN}{dA} = \left(\frac{dN}{dA}\right)_m
		{\rm exp} \left(-\frac{(\ln A - \ln \brakk{A})^2} {2 \ln \sigma_A}\right).
\end{equation}
This is valid for umbral areas $A_u$ larger than 
$A_{\rm min} = 1.5 \times 10^{-6} A_{1/2 \odot}$, where 
$A_{1/2 \odot} = 2 \pi R^2_\odot$ is the surface area of the 
visible solar hemisphere. Note that the total area of a sunspot $A_s$ 
is the sum of the umbral area $A_u$ and the much larger penumbral 
area $A_p$. Typical ratios of penumbral to umbral area vary between 
about 3 and 5 \cite{steinegger1990,brandt1990,beck1993}. In the following 
we assume that the penumbra is about 4 times larger than the 
umbra, so that $A_s = 5 A_u$.

The three free parameters that have to be determined by observations are
$(dN / dA)_m$, i.e.~the maximum value reached by the distribution, 
$\brakk{A}$, the mean sunspot umbral area and $\sigma_A$, a measure for the
width of the lognormal distribution. If all area measurements are 
taken into account, these 
parameters take on the values (in units of $10^{-6} A_{1/2 \odot}$) 
of $(dN / dA)_m = 9.4$, $\sigma_A=4.0$ and $\brakk{A} = 0.55$.
\scite{bogdan88} show that the same distribution can fit data from 
different cycles and that only variations in the
value of $(dN / dA)_m$ are statistically significant.

\section{Fits to the sunspot number distribution during different phases of
the solar cycle}
\label{sec:var_sun}

While \scite{bogdan88} show that it is {\em possible} to fit the 
sunspot umbral size spectrum with the same distribution, the best-fit
distributions are marginally different between sunspot maximum and minimum.
This difference between activity maximum and minimum is small, but a 
weighting towards larger spots could become important if we were
to extrapolate to higher activity levels such as the activity levels observed on
the stars listed in Tab.~\ref{tab:starlist}.

In the following sections we explore what variations in $\sigma_A$ 
and $\brakk{A}$ are consistent with the solar data and what these
would imply for the spot coverage of more active stars. As this 
depends on the number of free parameters of the fits, we separately 
investigate 2-degree fits with either $\brakk{A}$ or $\sigma_A$ fixed and
a 3-degree fit where $\brakk{A}$, $\sigma_A$ and $(dN / dA)_m$ are all 
allowed to vary simultaneously.

\subsection{Fits with fixed $\brakk{A}$}
The initial fits involved varying only $\sigma_A$ and $(dN / dA)_m$.
In the first instance we obtained a fit using the logarithmic form 
of Eq.~\ref{eq:dist}. Because of the logarithm, however, the errors are 
no longer normally distributed, so that there is no easy way to establish 
confidence limits. In order to obtain a reasonable estimate for the 
1-$\sigma$ deviation, we calculated the value of $\chi^2$ in the region 
surrounding the original fit (now using the non-logarithmic form, 
i.e. Eq.~\ref{eq:dist}). 
We hence have a 2-dimensional region given by $(dN / dA)_m$ and $\sigma_A$ 
where contours of constant $\chi^2$ describe ellipses. The constant-$\chi^2$
contours can be used to define a 1-$\sigma$
confidence region. For two degrees of freedom, the difference in $\chi^2$
between this region and the best fit is 2.3. 
The value for $(dN / dA)_m$, $\sigma_A$ and the deviation on $\sigma_A$
are given in Tab.~\ref{tab:sigvar}. In all cases the original ``logarithmic fit'' 
lies within the 1-$\sigma$ contour. All fits listed are for
$\brakk{A} =0.57$. 

\begin{figure}
	\centerline{\psfig{figure=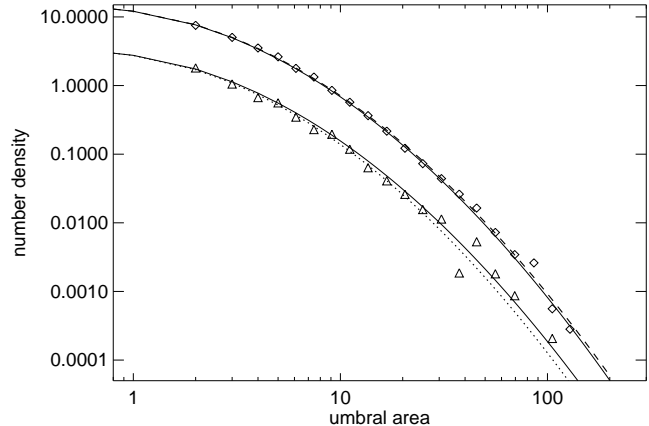,width=8.8cm}}
	\caption[]{Plot of the spot number density as a function of umbral 
	area (in units of 10$^{-6}$ solar hemisphere area) according
	to \protect\scite{bogdan88} and lognormal fits to the data. The 
	diamonds show the number density at solar maximum, the triangles
	the number distribution at solar minimum. The solid lines are the 
	best fit for all data sets (i.e. row 1 of Tab.~\protect{\ref{tab:sigvar}}), 
	multiplied by the maximum value of the distribution at maximum 
	and minimum. Also shown are the best fits to data taken during 
	solar maximum (dashed line) and solar minimum (dotted line, 
	rows 2 and 4 of Tab.~\protect{\ref{tab:sigvar}} respectively). 
	All fits are for varying $\sigma_A$ with $\brakk{A}$ fixed at 0.57. 
	}
	\label{fig:bogdan}
\end{figure}

\begin{table}
\caption[]{\label{tab:sigvar} Variations in $\sigma_A$ for a fixed $\brakk{A}$
of 0.57. The first column labels the cycle activity: ``all'' indicates that 
all available data were fitted; ``max'' and ``min'' are for data taken in the
years bracketing solar cycle maximum and minimum respectively; ``max+'' 
and ``min+'' also include the ascending respective descending phase 
of the cycle.  The second column gives the fit for $(dN/dA)_m$, 
the third and forth columns give the values for $\sigma_A$ and its 
standard deviation respectively. 
}
\begin{tabular}{lccc}
\hline 
cycle   & $(dN/dA)_m$ & $\sigma_A$ & $\Delta\sigma_A$	\\
\ \ [1]	&  [2]	& [3] 	& [4] \ \	       	\\
\hline 
all 	& \ 9.21 & 3.95	& 0.04	\\
max 	& 13.56	 & 4.02	& 0.05	\\
max+	& 11.51  & 4.04	& 0.04	\\
min	& \ 3.09 & 3.75	& 0.15	\\
min+	& \ 4.99 & 3.71	& 0.07	\\
\hline
\end{tabular}
\end{table}

Note that the fits to the data taken during solar minimum when
the number of sunspots on the disc is very small, are much less well
constrained than those taken during maximum. As a consequence, the 
1-$\sigma$ deviation at solar minimum is three times larger than at 
solar maximum. To achieve statistically more meaningful fits, we also 
considered a combined data set of solar minimum and the descending phase 
of the cycle (here labelled min+) and of solar maximum combined 
with the ascending phase of the cycle (labelled max+). While the best 
fits at solar minimum and maximum only deviate by about 2 $\Delta\sigma_A$, 
the best fits at phase min+ and max+ differ by 4 to 5 $\Delta\sigma_A$.
Some of the fits for solar minimum and maximum are
shown in Fig.~\ref{fig:bogdan} along with the spot number distribution at solar
minimum and solar maximum taken from \scite{bogdan88}. This shows that 
the distribution at solar minimum and maximum are indeed similar. 

\subsection{Fits with fixed $\sigma_A$}
The fits for varying values of $\brakk{A}$ and $(dN/dA)_m$ were obtained 
in a similar manner to the ones for varying $\sigma_A$. Here we fixed 
$\sigma_A$ to be 4.0 and again looked for constant $\chi^2$-difference contours 
at 2.3 for the 1-$\sigma$ confidence limits. We find that $\brakk{A}$ 
for the best fits varies between 0.49 and 0.58 for solar minimum and 
solar maximum respectively. The obtained values of $\brakk{A}$ are 
listed in Tab.~\ref{tab:Ameanvar}. 

\begin{figure}
	\centerline{\psfig{figure=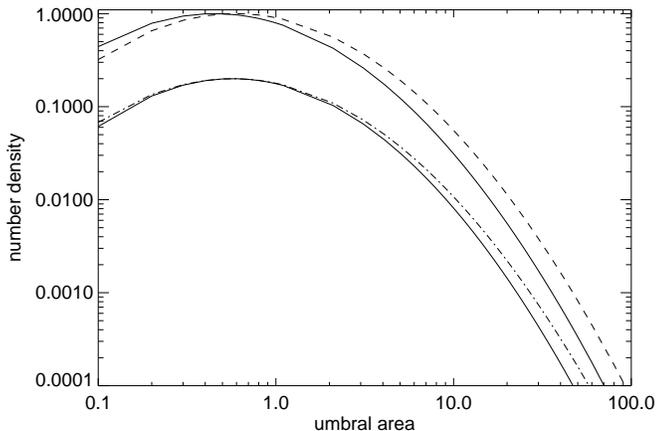,width=8.8cm}}
	\caption[]{The lognormal fits to the sunspot umbral distributions
	at solar minimum (solid lines) and maximum (broken lines). 
	The two uppermost lines show the fits for varying $\brakk{A}$, 
	the lower lines for varying $\sigma_A$. To bring out
	the small differences, we have taken the extreme values on the
	1-$\sigma$ contours. All fits have been normalised. For 
	clarity the varying-$\sigma_A$ fits have been offset.
}
\label{fig:lognorm}
\end{figure}

While varying $\sigma_A$ increases the width 
of the lognormal distribution, varying $\brakk{A}$ shifts the distribution 
towards larger $A_u$. Over the range of measured sunspot umbral sizes 
both methods yield equally good fits. The implications for small spots 
are, however, rather different. This is illustrated in 
Fig.~\ref{fig:lognorm} where the (normalised) lognormal distributions 
for fixed $\sigma_A$ and for fixed $\brakk{A}$ at solar minimum 
and maximum are compared. The solid lines show the fits at solar 
minimum. The dashed line for varying $\brakk{A}$ shows the shift towards 
higher umbral sizes, $A_u$. The dot-dashed line reveals the broader 
distribution obtained for a larger value of $\sigma_A$. Note that the 
umbral areas measured by \scite{bogdan88} range from about 2 to 100. In 
this range, varying $\sigma_A$ or $\brakk{A}$ gives an equivalent 
goodness-of-fit. Hence the difference in the fits mainly affects 
the number of pores, dark structures that are 
on average smaller than sunspots.  

\begin{table}
\caption[]{\label{tab:Ameanvar} Variations in $\brakk{A}$ for fixed $\sigma_A$.
The first column is as in the previous table and indicates the 
activity level. The second column gives $(dN/dA)_m$ and the third and 
forth columns give $\brakk{A}$ and its 1-$\sigma$ deviation.  
}
\begin{tabular}{lccc}
\hline 
cycle   & $(dN/dA)_m$ 	& $\brakk{A}$  & $\Delta\brakk{A}$ 	\\
\ \ [1] &  [2]  	& [3] 		& [4]  \\
\hline 
all     & \ 9.40 	& 0.56 		& 0.01	\\
max     & 13.37 	& 0.58 		& 0.01	\\
max$^+$ & 11.32 	& 0.58 		& 0.01	\\
min     & \ 3.45 	& 0.50 		& 0.05	\\
min$^+$ & \ 5.67 	& 0.49 		& 0.02	\\
\hline  
\end{tabular}     
\end{table}

\subsection{Variations in $\brakk{A}$ and $\sigma_A$}
If all three parameters are allowed to vary, the 1-$\sigma$ confidence region
is inside an ellipsoid whose surface has a value of $\chi^2$ that is higher
by 3.5 than the minimum value of $\chi^2$. However, the data are not sufficient
to constrain the three fit parameters very tightly and the 1-$\sigma$
deviations increase by more than a factor of 5 compared to the 
2-degree fits presented in the previous section. Furthermore, the relationship
between the $\brakk{A}$ and $\sigma_A$ parameters and the cycle characteristic
is no longer straightforward. Going from solar minimum to solar maximum 
in the previous sections implied larger values for $\brakk{A}$ or $\sigma_A$.
If $\brakk{A}$ and $\sigma_A$ are both allowed to vary, we tend to still get
larger values for $\brakk{A}$ (i.e.,~a shift towards larger mean spot sizes), 
though at the expense of $\sigma_A$ that now decreases (see columns 
3 and 4 of Tab.~\ref{tab:allvar}). This suggests, rather unexpectedly, 
a narrower distribution for higher levels of activity. 

When we look at the parameters within the 1-$\sigma$ confidence
regions, the picture becomes less clear, as there is overlap between 
the solar-maximum and solar-minimum parameters. Most parameters suggest
that the distribution at solar maximum is steeper as described above, 
but there are also some choices where the distribution at solar maximum 
is flatter, but shifted towards smaller-sized spots. This makes it rather 
difficult to pick scaling parameters (see Sec.~\ref{sec:All_extrapolate} for 
more details). 

\begin{table}
\caption[]{\label{tab:allvar} Variations in $\brakk{A}$ and $\sigma_A$. The
first column again lists the phase of the activity cycle.
The second column gives the best-fit values for $(dN/dA)_m$. Columns three and four
gives the best-fit value of $\sigma_A$ along with its deviation. The fifth and 
sixth columns list $\brakk{A}$ along with its deviation.
}
\begin{tabular}{lccccc}
\hline
cycle	& $(dN/dA)$ & $\sigma_A$ & $\Delta\sigma_A$	& $\brakk{A}$ 	& $\Delta\brakk{A}$ \\
\ \ [1] &  [2]  	& [3] 	& [4]			& [5] 		& [6] \\
\hline
all    & \ 9.4 		& 4.0 	& 0.3			& 0.56		& 0.10	\\
max    	& 12.4  	& 3.8 	& 0.3			& 0.64		& 0.09	\\
max+   	& 10.9  	& 3.9 	& 0.3			& 0.61		& 0.07	\\
min    & \ 4.2 		& 4.6 	& 3.1			& 0.39		& 0.19	\\
min+   & \ 6.4 		& 4.3 	& 0.8			& 0.43		& 0.10	\\ 
\hline
\end{tabular}

\end{table}

\section{Extrapolations to more active stars}
\label{sec:extpol}
\begin{figure}
        \centerline{\psfig{figure=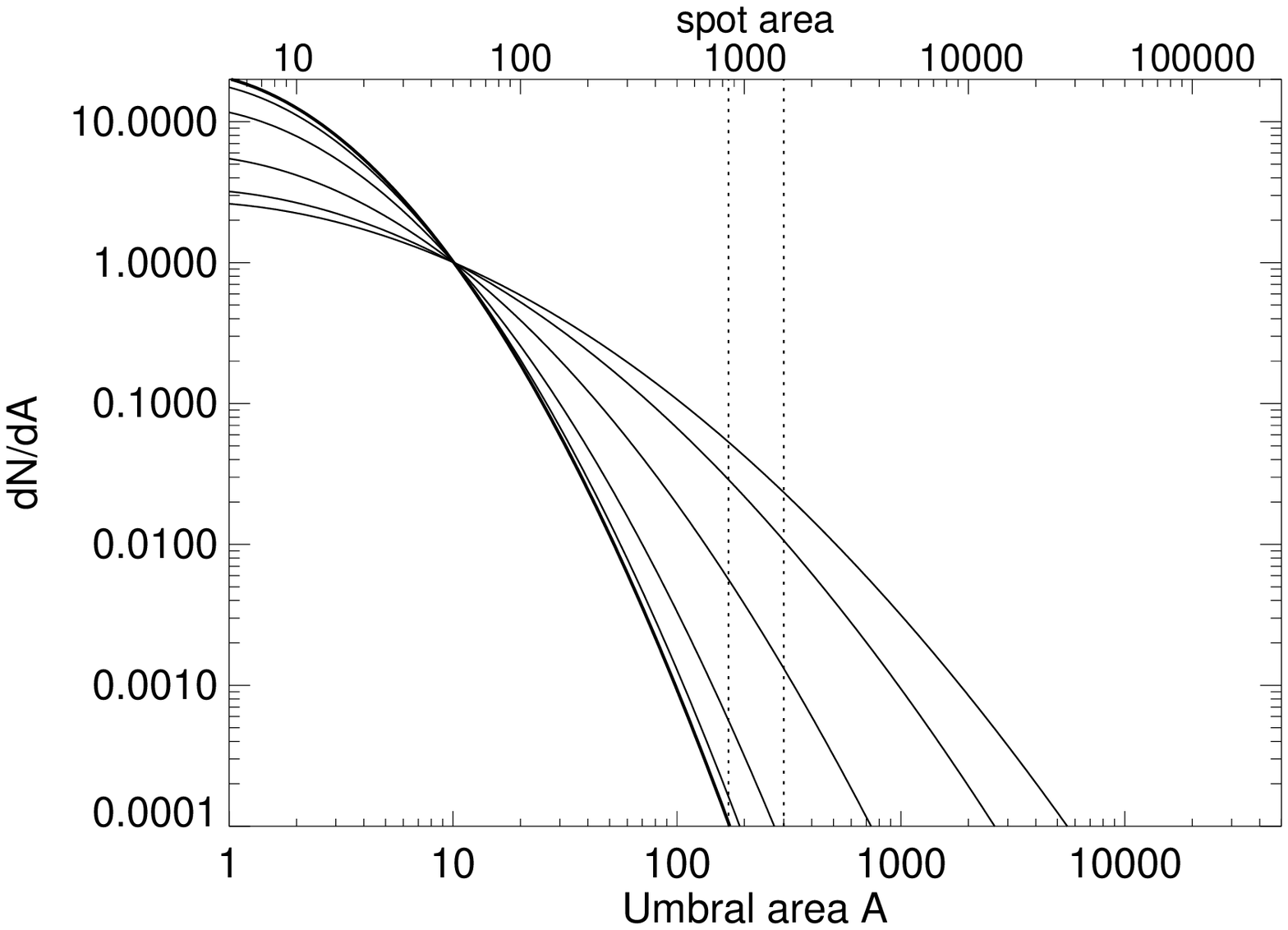,width=8.8cm}}
        \centerline{\psfig{figure=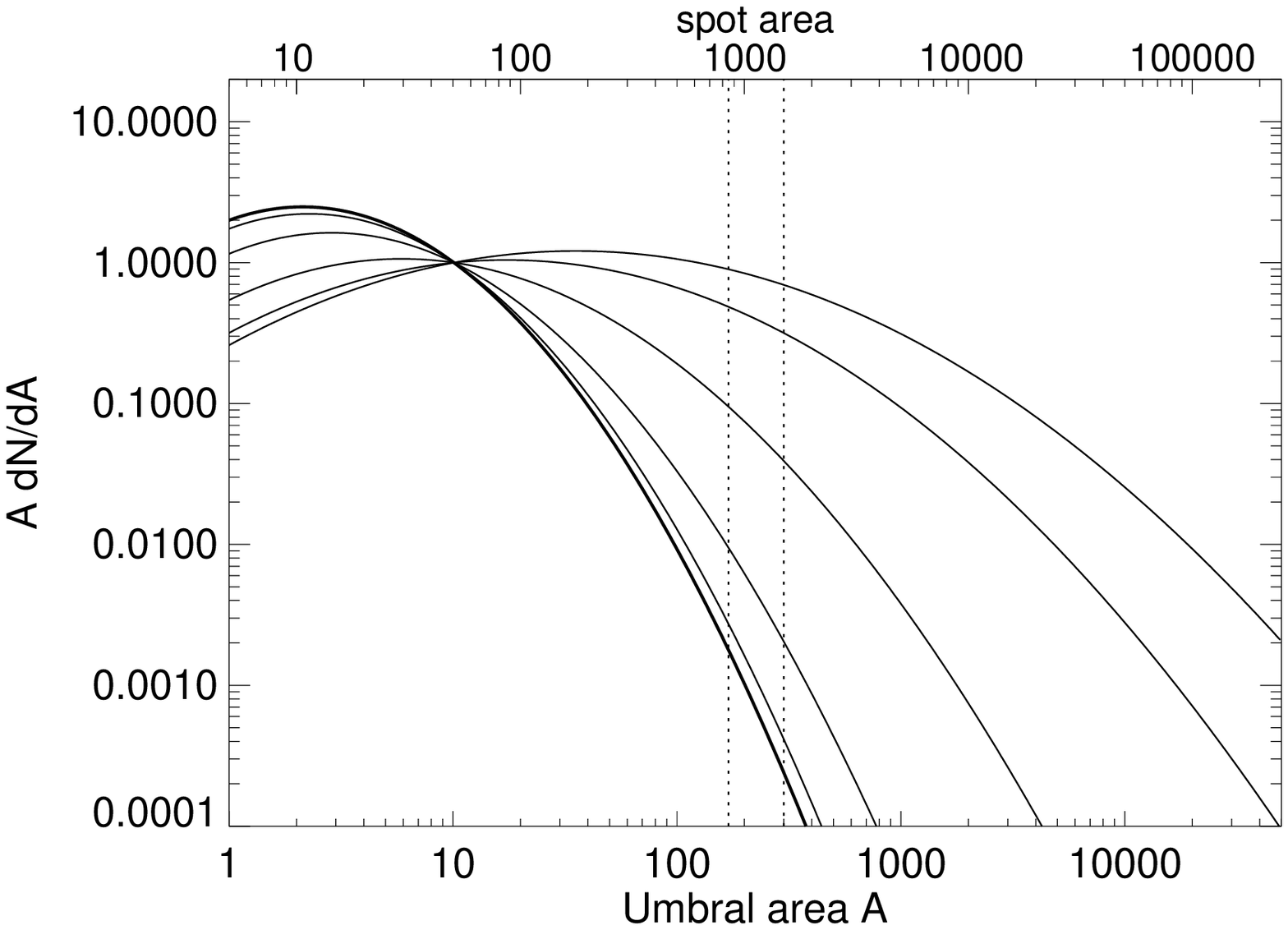,width=8.8cm}}
        \caption[]{Plots of number and size distribution $(dN/dA)$ (top)
        and $A (dN/dA)$ (bottom) of sunspot umbrae as a function 
	of spot umbral area $A_u$ in units of $10^{-6} A_{1/2 \odot}$. 
	It is assumed that $n_\sigma =1$, i.e., $\sigma_A$ is proportional
	to spot surface coverage.
	The distributions for the quiet Sun are represented by the thick lines. 
	The next line towards the right shows $(dN/dA)$ ($A (dN/dA)$ 
	for the lower plot) at solar maximum, while the lines further 
	to the right illustrate the extrapolations to more active stars, 
	up to a hypothetical star where about 75\% of one hemisphere is 
	covered by spots (see Tab.~\protect{\ref{tab:DIsigvar}} for a list of 
	parameters). The distributions have been normalised to 
	their values at an umbral area of $A=10 \times 10^{-6} A_{1/2 \odot}$. 
	The two dotted lines at 170$\times 10^{-6} A_{1/2 \odot}$
	and 300$\times 10^{-6} A_{1/2 \odot}$ indicate resolution 
	limits of Doppler imaging (see text for more detail).
        }
        \label{fig:dnda}
\end{figure}
\begin{figure}
        \centerline{\psfig{figure=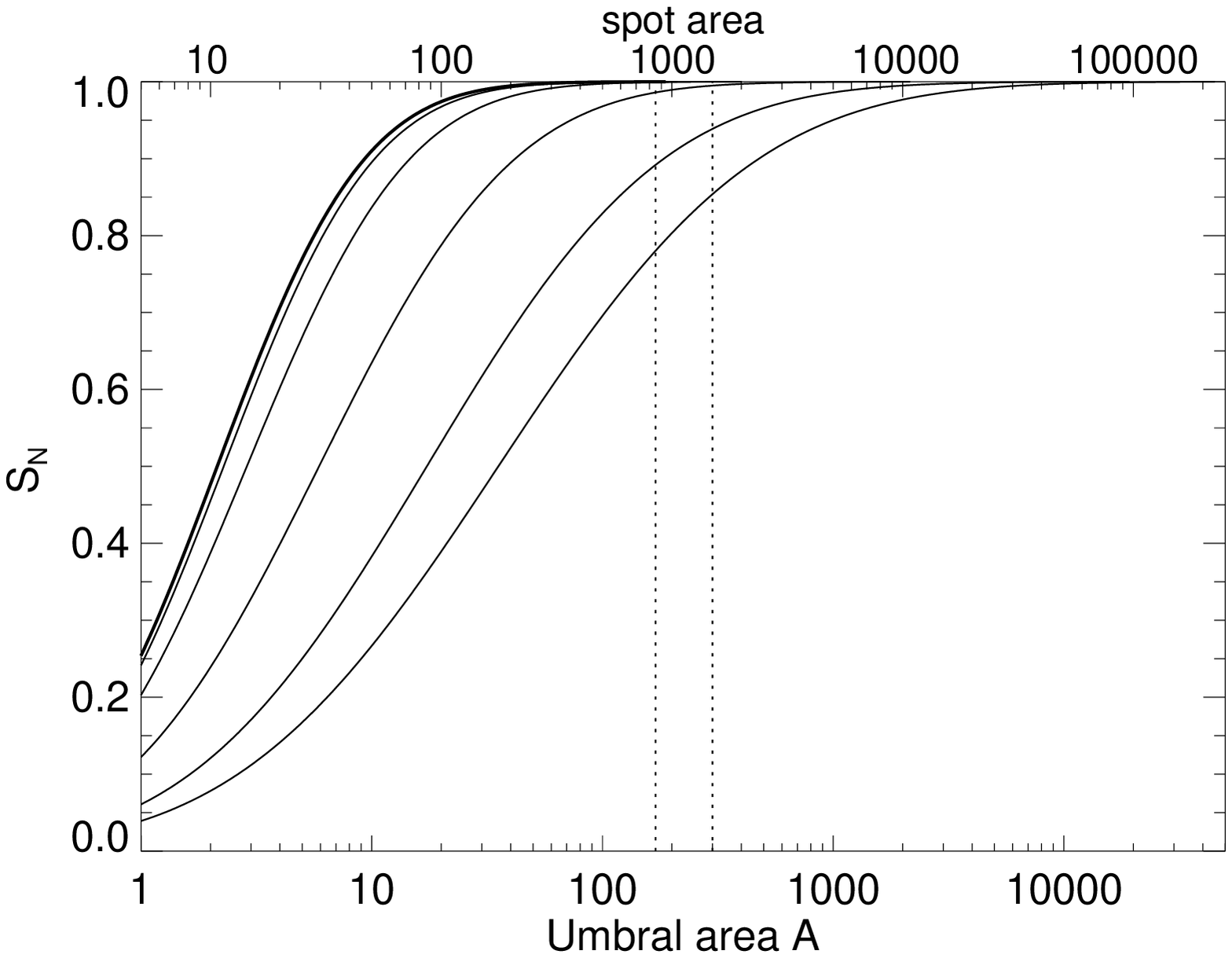,width=8.8cm}}
        \centerline{\psfig{figure=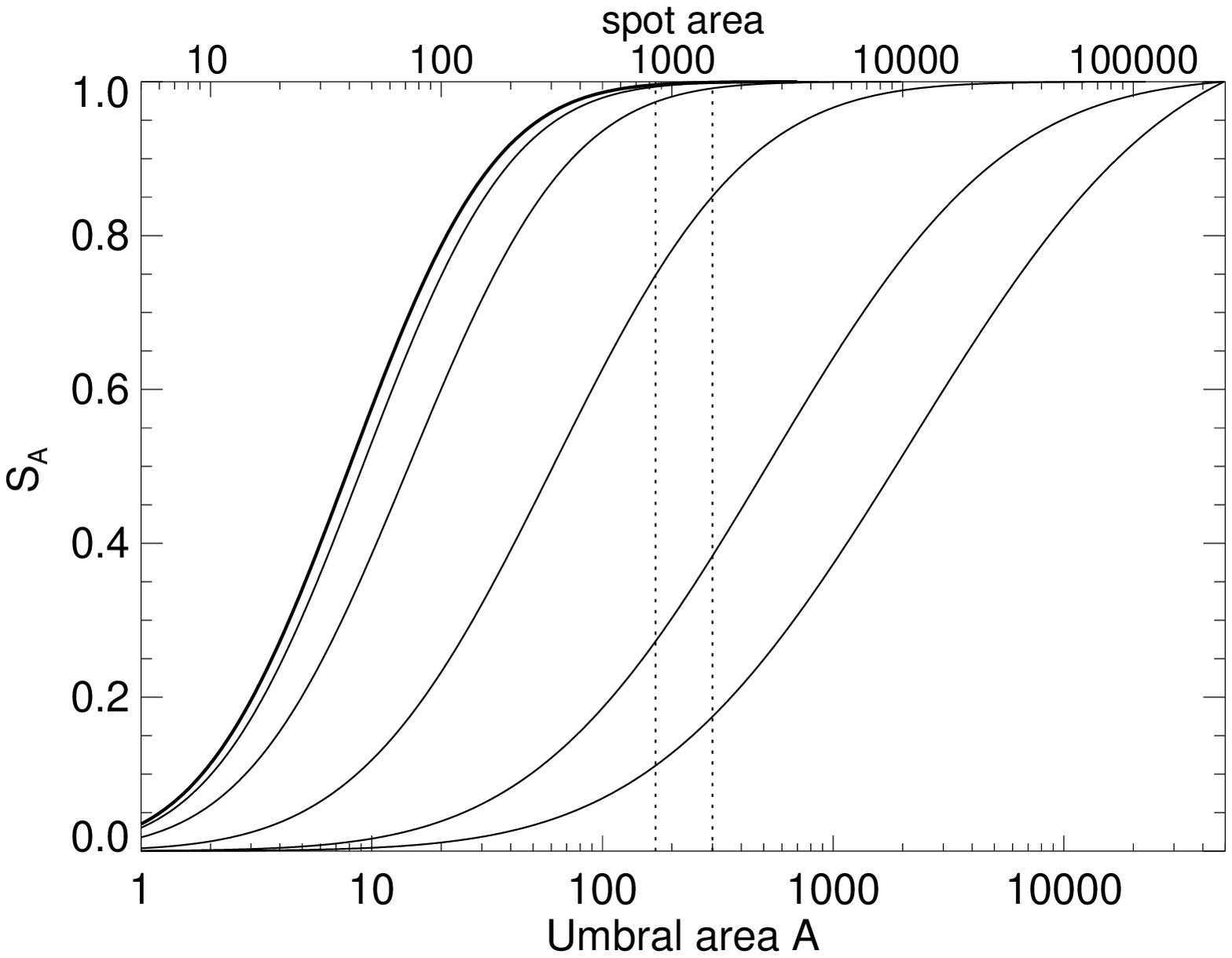,width=8.8cm}}
        \caption{Plots of the integral functions $S_N$ (top) and $S_A$ (bottom)
        as a function of umbral area $A$. The thicker lines show $S_N$ and $S_A$
        for the quiet Sun, while lines further to the right are for (linearly)
        increasing values of $\sigma_A$. See text for the definition of $S_N$
        and $S_A$ and Tab.~\ref{tab:DIsigvar} for a list of plot parameters.}
\label{fig:Sn}
\end{figure}

In addition to $(dN/dA)$ given by Eq.~\ref{eq:dist}, the following quantities
are of importance for the current analysis: $A (dN/dA)$, $S_N(A)$ and $S_A(A)$
(see also \pcite{solanki99armagh}). 
$S_N(A)$ and $S_A(A)$ are the integrals over $(dN/dA)$ and
$A (dN/dA)$, respectively, both being normalised to their maximum values:
\begin{eqnarray}
S_N(A) & = & \int_{A_{\rm min}}^A \frac{dN}{dA'} dA' \left/ 
	     \int_{A_{\rm min}}^{A_{\rm max}} \frac{dN}{dA'} dA' \right.    \nonumber\\
       & = & \int_{A_{\rm min}}^A dN / N_{\rm tot} , 
\label{eq:SN}
\end{eqnarray}
\begin{eqnarray}
S_A(A) & = & \int_{A_{\rm min}}^A A' \frac{dN}{dA'} dA' \left/ 
	     \int_{A_{\rm min}}^{A_{\rm max}} A' \frac{dN}{dA'} dA' \right.  \nonumber\\
       & = & \int_{A_{\rm min}}^A A' dN / A_{\rm tot} , 
\label{eq:SA}
\end{eqnarray}
$S_N(A)$ describes the relative contribution of spots with area between 
$A_{\rm min}$ and $A$ to the total number of spots, while $S_A(A)$ gives the
relative contribution of these spots to the total area covered by all spots 
on the solar or
stellar surface. The latter is hence the key quantity to compare with stellar
observations of different types such as molecular line strengths, which give
a measure of the total area covered by spots, $A_{\rm tot}$ (relative to the 
stellar surface area), and Doppler images, which provide information mainly
on the spots above a certain size. Example plots of $(dN/dA)$, $A (dN/dA)$, 
$S_N(A)$ and $S_A(A)$ for different activity levels are shown in 
Figs~\ref{fig:dnda} and \ref{fig:Sn}, and will be discussed in the following
section.

\subsection{Spot coverage for varying $\sigma_A$}
\label{sec:sigvar}
Having only two points to extrapolate from, we have in principle a large possible
range of scalings. In the following, we require that $\sigma_A$ scales with stellar 
activity as parameterised by the spot covering fraction, i.e., 
$\sigma_A = \sigma_A^0 + \Delta_\sigma (A_{\rm spot}/A_*)^{n_\sigma}$. 
The main open question concerns the choice of $n_\sigma$. Plots of solar magnetic 
activity proxies, such as Ca~{\sc ii}~H \& K or the 10.7~cm radio flux 
versus spot coverage show very large scatter and it is not clear how the spot
size distribution scales with magnetic activity. We therefore take the 
approach of choosing values of $n_\sigma$, carrying out the analysis and from a 
comparison with Doppler imaging and TiO results deciding whether our
choice is reasonable. Exponents $n_\sigma$ with values between 0.5 and 1.0 yield such 
results and will be discussed in the following. 

For the extrapolations shown here, $\brakk{A}$ is kept fixed at 0.57, while
we use the best-fit values of $\sigma_A=3.75$ and $\sigma_A=4.0$ at solar minimum
and maximum respectively. The values for $(dN/dA)_m$ at solar minimum and maximum
were adjusted so that the spot covering fraction at solar minimum and maximum were
0.03\% and 0.3\% respectively. Having preset the above parameters and picked 
a value for the exponent $n_\sigma$, we determine $\sigma_A^0$, the width of 
the lognormal distribution in the limit of zero spot coverage and $\Delta_\sigma$, 
the increase in the width with increasing activity. 

The solution for each activity level is then found by presetting $(dN/dA)_m$ and 
guessing a spot coverage $A_{\rm spot}/A_*$, and hence a new $\sigma_A$. 
Eq.~\ref{eq:dist} is then integrated to obtain a new spot coverage. 
This process is iterated until the calculated and input spot coverages agree.  
For exponents, $n_\sigma$, below about 0.7 this is straightforward 
with each next-higher value of $(dN/dA)_m$ yielding a solution with a 
higher spot coverage. For steeper exponents, there is a threshold value for 
$(dN/dA)_m$ beyond which no solutions can be found. But below the threshold value
there are generally two solutions, one for a low surface coverage and hence low 
$\sigma_A$, and one with a much higher surface coverage. This can be seen from 
Tab.~\ref{tab:DIsigvar} where for $n_\sigma=1$ the different parameters of the 
lognormal distribution are listed along with the calculated spot covering
fractions for different activity levels. The umbral-size distributions
for the same parameters are plotted in Fig.~\ref{fig:dnda}, 
the different lines correspond to the columns of Tab.~\ref{tab:DIsigvar}. 
The quiet-Sun behaviour is indicated by the thickest and left-most line.
The next line to the right is for solar maximum, with $\sigma_A$ increasing from
3.75 to 4.0 between solar minimum and maximum as outlined above. 
The two vertical dotted lines gives an indication of the currently achievable 
resolution with Doppler imaging. Only spots with areas larger than indicated by the 
dotted lines can be picked up.

\begin{table}
\caption[]{\label{tab:DIsigvar} Spot covering fractions and the proportion
of spots above the ``Doppler imaging threshold'' for $\sigma_A$
increasing linearly with stellar activity (stellar activity is parametrised with
the spot surface coverage). Rows 2 and 3 list the values for
$\sigma_A$ and $(dN/dA)_m$. The forth row gives the fractional spot
coverage of one hemisphere. The fifth and sixth rows give the fraction
of spots that can be seen on a Doppler map with 4$^{\circ}$
and 3$^{\circ}$ resolution respectively (note that row 5 (or 6) has to 
be multiplied with row 4 so as to obtain the fractional spot coverage 
that would be deduced from a Doppler image). 
}

	\begin{tabular}{lrrrrrrr}
	\hline
	solar: 		& min	& max	& 	& 	& 	& 	\\ \hline
        $\sigma_A$      & 3.75  & 4.00  & 5.04  & 10.4  & 30.3  & 62  \\
        $(dN/dA)_m$     & 4.5   & 38.5  & 106   & 106   & 41.2  & 20.8 \\
        $A_{\rm spot}/A_*$ & 0.0003 & 0.003 & 0.014 & 0.07 & 0.28 & 0.62        \\
        $A_{\rm DI}/A_{\rm tot}$
                        & \hspace{-3mm} 0.003 & 0.003 & 0.010 & 0.15  & 0.61    & 0.83  \\
                        & \hspace{-3mm} 0.006 & 0.008 & 0.027 & 0.25  & 0.73    & 0.89  \\ \hline
	\end{tabular}
\end{table}

If we ask what fraction of the stellar surface is covered by spots larger
than a given size, we have to consider the integral $S_A$. This integral
is shown for the parameters given in Tab.~\ref{tab:DIsigvar}
in the bottom plot of Fig.~\ref{fig:Sn}. The top graph of Fig.~\ref{fig:Sn} shows
the integral $S_N$. 
We have assumed that the current resolution limit for Doppler imaging is
around 3$^{\circ}$ or 4$^{\circ}$. For circular features this is equivalent
to effective surface areas of 340 and 600 $\times 10^{-6} A_{1/2 \odot}$.
The effective area does not correspond to the actual area of a starspot,
since the latter is probably composed of an umbra which produces a large contrast,
and a penumbra, which produces a smaller contrast per unit area. Hence the
effective area lies between the area of the starspot and that of its umbra.
To quantify this effective area better we have carried out some tests using 
a Doppler imaging code that is based on a spot-filling factor approach 
(see e.g.~\pcite{cameron94doppler}). They show that the relative contribution of a
spot umbra at 4500~K and a penumbra at 5400~K that is four times larger than
the umbra is similar\footnote{This is of course just a rough estimate
as the exact ratio depends on the particular line that is used for 
the mapping.}. Such a penumbra-to-umbra area ratio is typical of 
sunspots (see \scite{solanki2003spots} for an overview). We therefore assume that 
the umbral area of the smallest resolvable spot is 170$\times 10^{-6} A_{1/2 \odot}$,
respectively 300$\times 10^{-6} A_{1/2 \odot}$, as indicated by the
dotted lines in Figs~\ref{fig:dnda}, \ref{fig:Sn} and \ref{fig:dnda_Amean}.
This corresponds to total spot areas of 850$\times 10^{-6} A_{1/2 \odot}$
and 1500$\times 10^{-6} A_{1/2 \odot}$, i.e., spot diameters of
approximately 5$^{\circ}$ and 6$^{\circ}$.

\begin{table}
\caption[]{\label{tab:DIsigvar_root} Spot covering fractions and the proportion
of spots above the ``Doppler imaging threshold'' for $\sigma_A$ 
increasing as the square-root of stellar activity (see 
Tab.~\protect{\ref{tab:DIsigvar}} for explanations of the symbols).
}

        \begin{tabular}{lrrrrrrr}
        \hline
        solar:          & min   & max   &       &       &       &       \\ \hline
        $\sigma_A$      & 3.75  & 4.00  & 4.47  & 5.24  & 6.6   & 9.6  \\
        $(dN/dA)_m$     & 4.5   & 38.5  & 155   & 396   & 796   & 1393  \\
        $A_{\rm spot}/A_*$ & 0.0003 & 0.003 & 0.015 & 0.06 & 0.19 & 0.78        \\
        $A_{\rm DI}/A_{\rm tot}$
                        & \hspace{-3mm} 0.004 & 0.004 & 0.005 & 0.01  & 0.04    & 0.12  \\ 
                        & \hspace{-3mm} 0.007 & 0.009 & 0.015 & 0.03  & 0.08    & 0.22  \\ \hline
        \end{tabular}
\end{table}

The percentage of spots that will be picked up by Doppler imaging is listed in 
the bottom two rows of Tab.~\ref{tab:DIsigvar} for $\sigma_A$ increasing 
linearly with the spot surface 
coverage, i.e. $n_{\sigma}=1$. The individual columns correspond to the lines 
drawn in  Fig.~\ref{fig:Sn}. It turns out that for very large $\sigma_A$ characterising
wide lognormal spot distributions a substantial fraction of spots that are present
on the stellar surface are indeed seen on the Doppler maps. We note that this 
is not the case for slower scaling laws ($n_{\sigma} < 0.7$), where even for 
very large covering fractions less than half the spots are picked up on the Doppler maps. 
Tab.~\ref{tab:DIsigvar_root} lists the spot coverage and pick-up fraction for 
$n_{\sigma} =0.5$. 

\begin{figure}
        \centerline{\psfig{figure=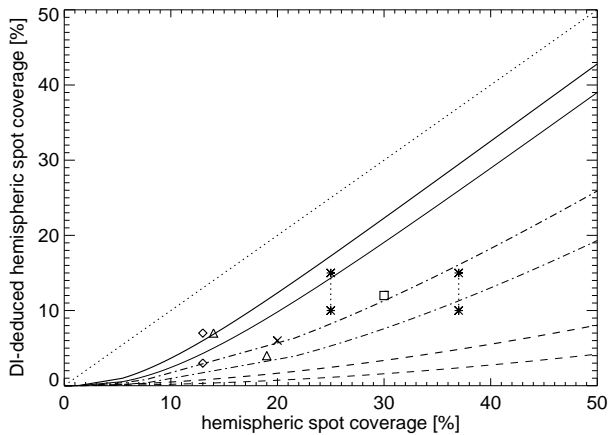,width=8.8cm}}
        \caption[]{Plots of the surface coverage deduced from Doppler 
	imaging as a function of actual surface coverage.
	The solid lines are for $\sigma_A$ increasing linearly with the 
	surface-spot coverage ($n_\sigma=1$), the dot-dashed lines are for
	$n_\sigma=0.75$ and the dashed lines for $n_\sigma=0.5$. The lower-lying 
	thin lines are for a resolution threshold of 4$^\circ$, while the upper heavier 
	lines are for a resolution threshold of 3$^\circ$. The dotted 
	line is plotted as a help to mark out a situation where all spots 
	present on the star would also appear on the Doppler map. 
	Also shown are some data points from Tab.~\protect{\ref{tab:starcover}}
	where simultaneous Doppler images and TiO filling factor estimates 
	are available. These TiO estimates were used to represent the spot coverage
	of one hemisphere. The points plotted are for II~Peg (stars), EI~Eri (diamonds), 
	$\sigma$~Gem (triangles), DM~UMa (square) and HD~199178 (cross). 
        }
\label{fig:sigvar}
\end{figure}

Different scaling laws are also contrasted in Fig.~\ref{fig:sigvar} 
where the spot area seen on Doppler images is plotted against the actual spot area.
The solid lines are for $n_\sigma=1$, the dashed lines for $n_\sigma=0.5$
and the dot-dashed lines for the intermediate case of $n_\sigma=0.75$. The symbols
denote data points from Tab.~\ref{tab:starcover} where simultaneous Doppler maps
and TiO filling factors were available. 
The stark difference between the square-root and linear scaling laws is due to 
the pronounced shape-change of the lognormal distribution for a linear increase. 
This results in a greater fraction of large spots compared to ``average'' spots. 
The {\em number} of smaller spots of course also increases, but this is negligible 
because of their small contribution to the total spot area. 
In the square-root scaling, the increase in the total spot area is achieved mainly 
by increasing the height of the distribution rather than its width as indicated by 
the ever increasing $(dN/dA)_m$ in Tab.~\ref{tab:DIsigvar_root}. 
This produces a much slower change in the fraction of large to average-size spots
so that only a very limited number of spots fall above the detection threshold.
It is clear from Tabs~\ref{tab:DIsigvar} and \ref{tab:DIsigvar_root} and from 
Fig.~\ref{fig:sigvar} that the curves for $n_\sigma=1$ and 0.5 lie sufficiently 
far apart to encompass most of the data points. In fact, we find that 
most data points can be accommodated with $n_{\sigma}$ between 0.75 and 1.

\subsection{Spot coverage for varying $\brakk{A}$}
\label{sec:Avar}
\begin{figure}
        \centerline{\psfig{figure=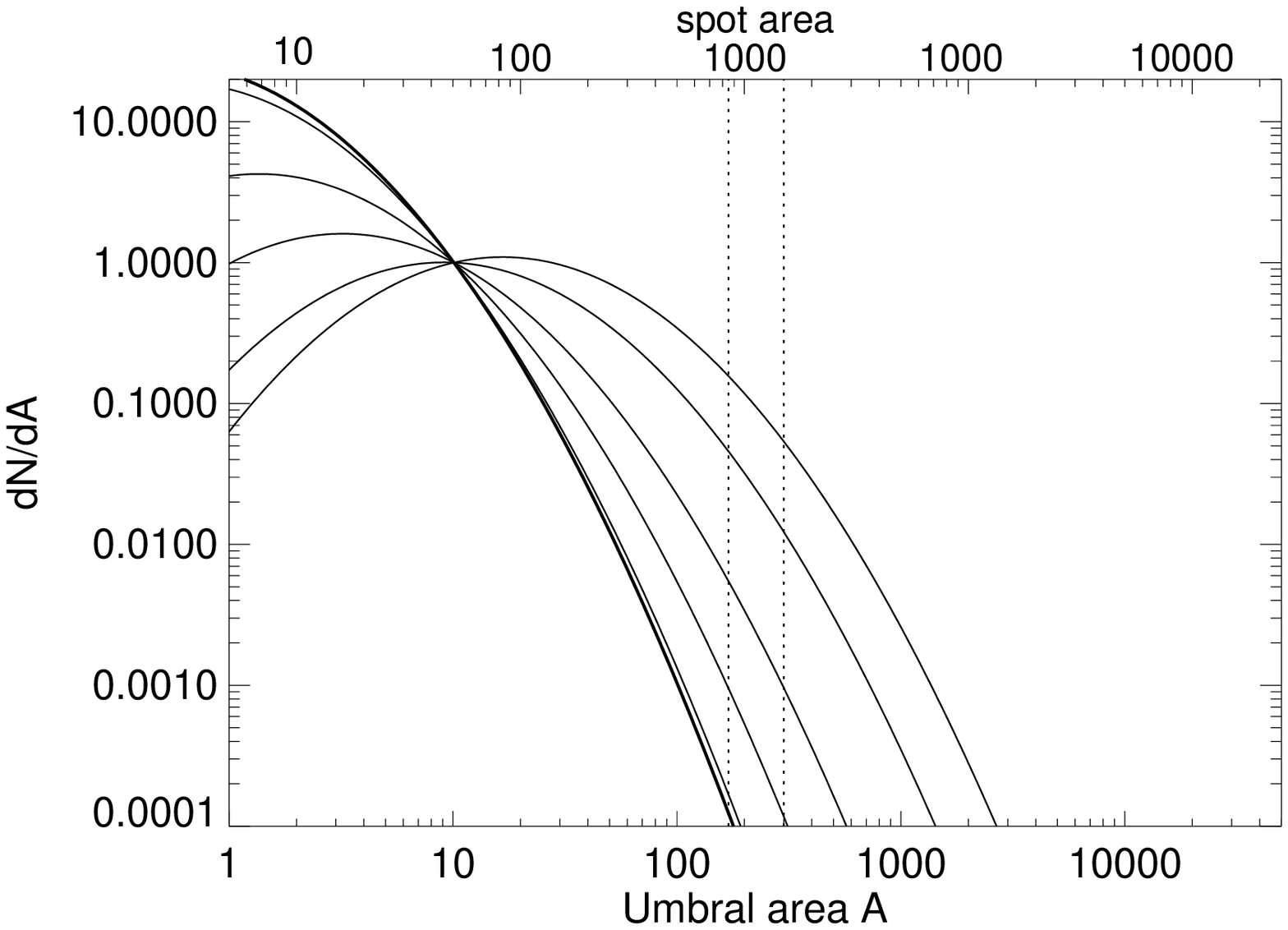,width=8.8cm}}
        \centerline{\psfig{figure=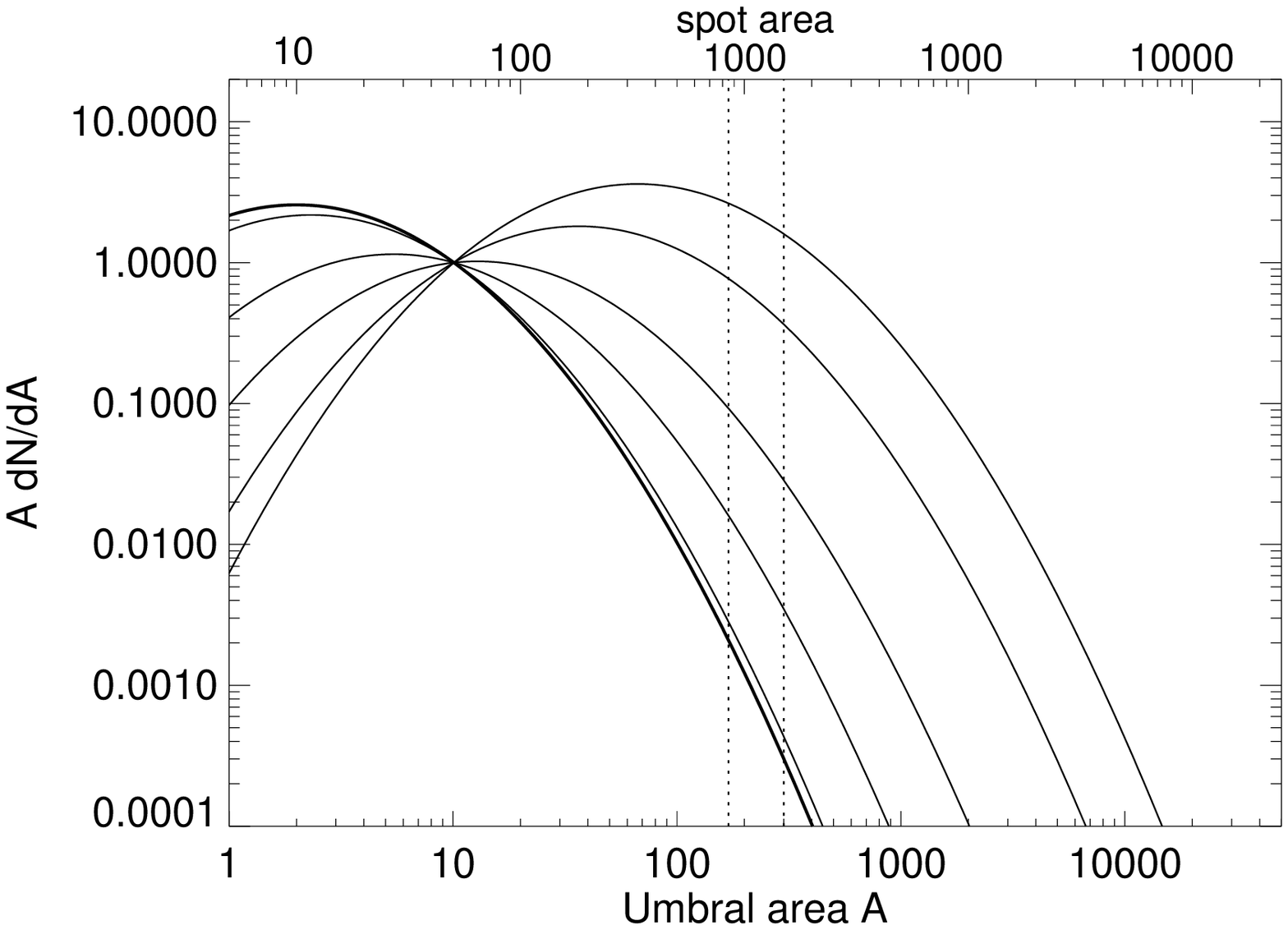,width=8.8cm}}
        \caption[]{Plots of number and size distribution $(dN/dA)$ (top)
        and $A (dN/dA)$ (bottom) of sunspot umbrae as a function
        of spot umbral area. The distributions for the quiet Sun are represented 
	by the thick lines. The next line towards the right corresponds to 
        solar maximum, while the lines further to the right illustrate the 
	extrapolations to more active stars where $\brakk{A}$ increases 
	linearly with the total spot coverage. See Fig.~\protect{\ref{fig:dnda}} 
	for a comparison with extrapolations in $\sigma_A$ and 
	Tab.~\protect{\ref{tab:Avar}} for a list of the parameters used for 
	this plot.
        }
        \label{fig:dnda_Amean}
\end{figure}

We now keep $\sigma_A$ fixed and extrapolate by allowing $\brakk{A}$
to vary. The assumption in this case is that as we move from solar 
minimum through solar maximum and on to more 
active stars, the mean spot size, $\brakk{A}$, increases, thereby shifting the 
distribution towards higher values of $A$. The mean spot size increases 
with spot coverage according to 
$\brakk{A} = \brakk{A}^0 + \Delta_A (A_{\rm spot}/A_*)^{n_A}$. Again, 
the value for $n_A$ is the biggest unknown. Here we illustrate the results
for the same values as for $n_\sigma$, i.e. we assume $n_A=0.5$, 0.75 and 1.0.
While the fits to the 
solar minimum and maximum distribution with either varying $\sigma_A$ 
or $\brakk{A}$ are very similar (see Fig.~\ref{fig:lognorm}), the 
distributions differ strongly for more active stars. This is illustrated
in Fig.~\ref{fig:dnda_Amean} where the number and size distributions for 
increasing $\brakk{A}$ have been plotted. Note the much smaller number of large 
spots for the ``shifted'' distribution compared to the ``widened'' distribution
shown in Fig.~\ref{fig:dnda}.

\begin{table}
\caption[]{\label{tab:Avar} Spot covering fractions and the proportion
of spots above the ``Doppler imaging threshold'' for $\brakk{A}$
increasing linearly with stellar activity, as parametrised with
the spot surface coverage. See Tab.~\protect{\ref{tab:DIsigvar}} for 
a description of the symbols.
}

        \begin{tabular}{lrrrrrrr}
        \hline
        solar:          & min   & max   &       &       &       &       \\ \hline
        $\brakk{A}$     & 0.50  & 0.58  & 1.36  & 3.2   & 9.1   & 16.7  \\
        $(dN/dA)_m$     & 5     & 37    & 66    & 37    & 15    & 8   \\
        $A_{\rm spot}/A_*$ & 0.0003 & 0.003 & 0.03 & 0.09 & 0.29 & 0.54        \\
        $A_{\rm DI}/A_{\rm tot}$
          & \hspace{-3mm} 0.001 & 0.003 & 0.015 & 0.07  & 0.27  & 0.46  \\
          & \hspace{-3mm} 0.005 & 0.009 & 0.042 & 0.15  & 0.45  & 0.65  \\ \hline
        \end{tabular}
\end{table}

\begin{figure}
        \centerline{\psfig{figure=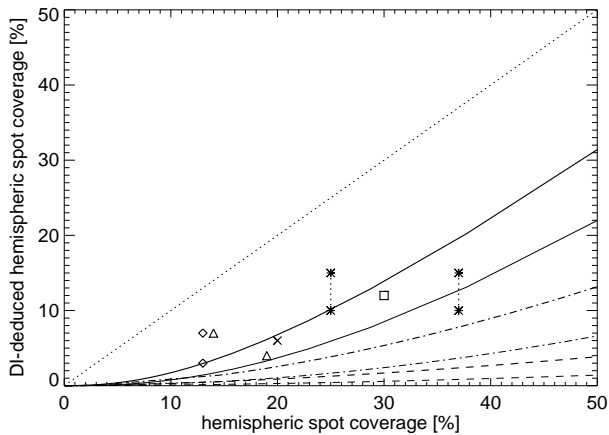,width=8.8cm}}
        \caption[]{
	Plots of the surface coverage deduced from Doppler
        imaging as a function of actual surface coverage 
	for scaling laws with $\brakk{A} = \brakk{A}^0 + \Delta_A (A_{\rm spot}/A_*)^{n_A}$.
        The solid, dot-dashed and dashed lines are for exponents, $n_A$, of
	1.0, 0.75 and 0.5 respectably. The lower-lying
        thin lines are for a resolution threshold of 4$^\circ$, the 
	upper heavier lines for a resolution threshold of 3$^\circ$. 
	The symbols are data points from Tab.~\protect{\ref{tab:starcover}}
	(see caption of Fig.~\protect{\ref{fig:sigvar}} for more detail). 
        }
\label{fig:Avar}
\end{figure}

The extrapolations for $\brakk{A}$ were carried out in a similar manner as
described in the previous section, with 
$\brakk{A} = \brakk{A}^0 + \Delta_A (A_{\rm spot}/A_*)^{n_A}$.
The fixed values are $\sigma_A=4.0$ for all fits, $\brakk{A} = 0.50$ 
at solar minimum and $\brakk{A} = 0.58$ at solar maximum. 
Some example parameters and results are listed in Tab.~\ref{tab:Avar}.
We find that the ``shift'' of the lognormal distribution towards larger $A$
is less efficient in creating spots that are large enough to be picked up on 
Doppler maps. This can be seen by comparing Figs~\ref{fig:sigvar} and 
\ref{fig:Avar}, where the spot covering fractions that would be seen on 
a typical Doppler map are plotted against the actual spot covering fraction. 
On both plots the thicker lines are for a resolution of 3$^{\circ}$, while 
the thinner (and lower-lying) lines are for a resolution of 4$^{\circ}$.
Fig.~\ref{fig:Avar} also shows the spot covering fractions that would 
be picked up for $n_A = 0.75$ (dot-dashed lines) and $n_A = 0.5$ (dashed lines). 
Note that the pick-up rates for $n_A = 0.5$ are almost one order
of magnitude smaller than those predicted for a linear increase in $\brakk{A}$. 
Our calculations suggest that if, indeed, the lognormal distribution scales with 
mean spot size, then some spot clumping is needed to explain the 
relatively high pick-up rates on the Doppler maps. Starting from solar 
observations as well as from observations of active longitudes
on rapidly rotating stars, some amount of spot clumping is in fact expected. 

\subsection{Spot covering for varying $\sigma_A$ and $\brakk{A}$}
\label{sec:All_extrapolate}
When both, $\sigma_A$ and $\brakk{A}$, are allowed to vary between solar minimum
and maximum, we get the somewhat curious situation that, for the best fits, only 
the value for $\brakk{A}$ increases between minimum and maximum, while the value 
of $\sigma_A$ decreases (see Tab.~\ref{tab:allvar}). 
Extrapolating from there we can reproduce the results found in 
Sects~\ref{sec:sigvar} and \ref{sec:Avar}, i.e., obtain curves similar to those 
plotted in Figs~\ref{fig:sigvar} and \ref{fig:Avar}. Since there are now two 
free parameters ($n_\sigma$ and $n_A$) compared to only one in the previous 
sections, this is not surprising and we do not learn anything new. Therefore 
we refrain from discussing these extrapolations in detail. 

\section{Discussion and conclusions}
\label{sec:conclusion}

We consider some of the consequences if spot sizes follow a lognormal 
distribution as on the sun. \scite{bogdan88} proposed that the 
passage of magnetic flux tubes from the dynamo to the stellar
surface through the turbulent convection zone leads to a certain 
fragmentation of the flux, producing a lognormal distribution of 
umbral areas (flux-tube cross-sections). 
In this picture a significant fraction of the total starspot 
area of any star is in the form of starspots below the resolution limit 
of Doppler images. The sunspot-size distribution is consistent with 
a shift towards larger spots at solar activity maximum. We have used the 
possible range in the solar parameters to extrapolate to higher 
activity levels in different ways and have compared the resulting 
fractional spot areas that can be resolved by Doppler imaging. This
has been done assuming  that the starspots are randomly distributed on the
stellar surface, which implies that most starspots resolved by Doppler
imaging are single spots. Hence we assume that starspots
do not clump. If they clump together as in solar 
active regions then Doppler images detect a larger fraction of the 
starspots than suggested by our analysis.

Recent calculations of \scite{berdyugina2002potsdam} suggest that 
at least for the RS\,CVn star II~Peg, spectral synthesis of the 
TiO band reproduces the observations of these molecular lines without 
requiring starspots in addition to those present on the corresponding Doppler 
image.  If starspot distributions on RSCVn stars are indeed log-normal and follow 
a sun-like pattern, this result would suggest that the starspots are tightly 
clumped and that the starspots resolved by Doppler imaging are actually 
conglomerates of smaller 
spots. The low photospheric temperature of II~Peg (Table~\ref{tab:starlist}), 
however, means that CN lines blending the analysed TiO band are also 
present in the spectrum of the immaculate star (Berdyugina, priv.~comm.). 
Due to uncertainties in the exact temperature there are also some 
uncertainties in the above result and further such calculations for hotter stars
would be of great interest. 

If larger spots increase in number more rapidly 
than smaller spots as stars become more active, we also expect starspots 
as a whole to be more numerous relative to smaller magnetic flux tubes, i.e., 
bright magnetic elements.  The ``switch-over'' between activity-bright and 
activity-dark stars seen with increasing activity level 
\nocite{radick87,radick90,radick98,lockwood92} (Radick et al.~1987; 1990; 1998; 
Lockwood et al.~1992) fits well into this, as does 
the strong increase of spot area relative to facular area from solar activity 
minimum to maximum \cite{chapman97}. The fact that model calculations based
on extrapolations from solar values do reproduce the switch-over at about 
the correct activity level \cite{knaack98diplom}, supports our general 
approach of extrapolating the size distribution of spots from the 
Sun to more active stars. 

\section*{ACKNOWLEDGMENTS}
The authors would like to thank T.~Bogdan for digging out his notes and 
providing us with his data on the spot-size distributions. 
\bibliographystyle{/home/ycu/Tex/MNRAS/mn}
\bibliography{/home/ycu/Tex/Bib/iau_journals,/home/ycu/Tex/Bib/master,/home/ycu/Tex/Bib/myown,/home/ycu/Tex/Bib/accrefs,/home/ycu/Tex/Bib/spotsize,/home/ycu/Tex/Bib/limb,/home/ycu/Tex/Bib/solar}

\begin{thebibliography}{{{Washuettl}, {Strassmeier} \&
  {Collier-Cameron}}{2001}}

\bibitem[\protect\citefmt{{Beck} \& {Chapman}}{1993}]{beck1993}
{Beck}~J.~G., {Chapman}~G.~A., 1993, Solar~Phys., 146, 49

\bibitem[\protect\citefmt{Berdyugina \&
  Usoskin}{2003}]{berdyugina2003longitude}
Berdyugina~S.~V., Usoskin~I.~G., 2003, A\&A, submitted

\bibitem[\protect\citefmt{{Berdyugina} {\rm
  et~al.}}{1998}]{berdyugina98iipeg_di}
{Berdyugina}~S.~V., {Berdyugin}~A.~V., {Ilyin}~I., {Tuominen}~I., 1998, A\&A,
  340, 437

\bibitem[\protect\citefmt{Berdyugina}{2002}]{berdyugina2002potsdam}
Berdyugina~S.~V., 2002, Astron.~Nachr., 323, 192

\bibitem[\protect\citefmt{Bogdan {\rm et~al.}}{1988}]{bogdan88}
Bogdan~T.~J., Gilman~P.~A., Lerche~I., Howard~R., 1988, ApJ, 327, 451

\bibitem[\protect\citefmt{{Brandt}, {Schmidt} \&
  {Steinegger}}{1990}]{brandt1990}
{Brandt}~P.~N., {Schmidt}~W., {Steinegger}~M., 1990, Solar~Phys., 129, 191

\bibitem[\protect\citefmt{Chapman, Cookson \& Dobias}{1997}]{chapman97}
Chapman~G.~A., Cookson~A.~M., Dobias~J.~J., 1997, ApJ, 482, 541

\bibitem[\protect\citefmt{Collier~Cameron \& Unruh}{1994}]{cameron94doppler}
Collier~Cameron~A., Unruh~Y.~C., 1994, MNRAS, 269, 814

\bibitem[\protect\citefmt{{Dempsey} {\rm et~al.}}{1992}]{dempsey1992}
{Dempsey}~R.~C., {Bopp}~B.~W., {Strassmeier}~K.~G., {Granados}~A.~F.,
  {Henry}~G.~W., {Hall}~D.~S., 1992, ApJ, 392, 187

\bibitem[\protect\citefmt{{Hatzes}}{1993}]{hatzes93siggem}
{Hatzes}~A.~P., 1993, ApJ, 410, 777

\bibitem[\protect\citefmt{Hatzes}{1995}]{hatzes95}
Hatzes~A.~P., 1995, AJ, 109, 350

\bibitem[\protect\citefmt{{Henry} {\rm et~al.}}{1995}]{henry95rscvn}
{Henry}~G.~W., {Eaton}~J.~A., {Hamer}~J., {Hall}~D.~S., 1995, ApJS, 97, 513

\bibitem[\protect\citefmt{Knaack}{1998}]{knaack98diplom}
Knaack~R., 1998.
\newblock {\it Diplomarbeit}, {\it Institute of Astronomy, ETH, Z\"urich}

\bibitem[\protect\citefmt{Lockwood {\rm et~al.}}{1992}]{lockwood92}
Lockwood~G.~W., Skiff~B.~A., Baliunas~S.~L., Radick~R.~R., 1992, Nature, 360,
  653

\bibitem[\protect\citefmt{{Neff}, {O'Neal} \& {Saar}}{1995}]{neff95iipeg_tio}
{Neff}~J.~E., {O'Neal}~D., {Saar}~S.~H., 1995, ApJ, 452, 879

\bibitem[\protect\citefmt{{O'Neal}, {Neff} \& {Saar}}{1998}]{oneal98tio}
{O'Neal}~D., {Neff}~J.~E., {Saar}~S.~H., 1998, ApJ, 507, 919

\bibitem[\protect\citefmt{{O'Neal}, {Saar} \& {Neff}}{1996}]{oneal96tio}
{O'Neal}~D., {Saar}~S.~H., {Neff}~J.~E., 1996, ApJ, 463, 766

\bibitem[\protect\citefmt{{Padmakar} \& {Pandey}}{1999}]{padmakar99}
{Padmakar}, {Pandey}~S.~K., 1999, A\&AS, 138, 203

\bibitem[\protect\citefmt{Petrovay}{2001}]{petrovay2001}
Petrovay~K., 2001, in Wilson~A., ed, Proceedings of the 1st Solar \& Space
  Weather Euroconference: The Solar Cycle and Terrestrial Climate.
\newblock ESA SP-4, p.~63

\bibitem[\protect\citefmt{Radick {\rm et~al.}}{1987}]{radick87}
Radick~R.~R., Thompson~D.~T., Lockwood~G.~W., Duncan~D.~K., Baggett~W.~E.,
  1987, ApJ, 321, 459

\bibitem[\protect\citefmt{{Radick} {\rm et~al.}}{1998}]{radick98}
{Radick}~R.~R., {Lockwood}~G.~W., {Skiff}~B.~A., {Baliunas}~S.~L., 1998, ApJS,
  118, 239

\bibitem[\protect\citefmt{Radick, Skiff \& Lockwood}{1990}]{radick90}
Radick~R.~R., Skiff~B.~A., Lockwood~G.~W., 1990, ApJ, 353, 524

\bibitem[\protect\citefmt{{Rodon{\`o}} {\rm et~al.}}{2000}]{rodono2000iipeg}
{Rodon{\`o}}~M., {Messina}~S., {Lanza}~A.~F., {Cutispoto}~G., {Teriaca}~L.,
  2000, A\&A, 358, 624

\bibitem[\protect\citefmt{{Saar} \& {Neff}}{1990}]{neff1990tio}
{Saar}~S.~H., {Neff}~J.~E., 1990, in ASP~Conf.~Ser.~9: Cool Stars, Stellar
  Systems, and the Sun.
\newblock p.~171

\bibitem[\protect\citefmt{Saar}{1986}]{saar88}
Saar~S.~H., 1986, ApJ, 324, 441

\bibitem[\protect\citefmt{{Schrijver} \& {Title}}{2001}]{schrijver2001}
{Schrijver}~C.~J., {Title}~A.~M., 2001, ApJ, 551, 1099

\bibitem[\protect\citefmt{Sch\"ussler \& Schmitt}{2002}]{schussler2002}
Sch\"ussler~M., Schmitt~D., 2002, in Solar Variability and its Effect on the
  Earth's Atmospheric and Climate System.
\newblock American Geophysical Union, p.~3, in press

\bibitem[\protect\citefmt{Sch\"{u}ssler \& Solanki}{1992}]{schuessler92}
Sch\"{u}ssler~M., Solanki~S.~K., 1992, A\&A, 264, L13

\bibitem[\protect\citefmt{{Sch\"ussler} {\rm et~al.}}{1996}]{schussler96buoy}
{Sch\"ussler}~M., Caligari~P., Ferriz-Mas~A., Solanki~S.~K., Stix~M., 1996,
  A\&A, 314, 503

\bibitem[\protect\citefmt{{Solanki}}{1992}]{solanki92cs7}
{Solanki}~S., 1992, in ASP Conf.~Ser.~26: Cool Stars, Stellar Systems, and the
  Sun.
\newblock p.~211

\bibitem[\protect\citefmt{{Solanki}}{1999}]{solanki99armagh}
{Solanki}~S.~K., 1999, in Butler~C.~J., Doyle~J.~G., eds, ASP Conf. Ser. 158:
  Solar and Stellar Activity: Similarities and Differences.
\newblock p.~109

\bibitem[\protect\citefmt{Solanki}{2003}]{solanki2003spots}
Solanki~S.~K., 2003, A\&AR, 11, 153

\bibitem[\protect\citefmt{{Steinegger} {\rm et~al.}}{1990}]{steinegger1990}
{Steinegger}~M., {Brandt}~P.~N., {Schmidt}~W., {Pap}~J., 1990, Ap\&SS, 170, 127

\bibitem[\protect\citefmt{{Strassmeier} {\rm
  et~al.}}{1999}]{strassmeier99hd199178}
{Strassmeier}~K.~G., {Lupinek}~S., {Dempsey}~R.~C., {Rice}~J.~B., 1999, A\&A,
  347, 212

\bibitem[\protect\citefmt{Unruh, Collier~Cameron \&
  Cutispoto}{1995}]{unruh95doppler}
Unruh~Y.~C., Collier~Cameron~A., Cutispoto~G., 1995, MNRAS, 277, 1145

\bibitem[\protect\citefmt{{Washuettl}, {Strassmeier} \&
  {Collier-Cameron}}{1998}]{washuettl98cs10}
{Washuettl}~A., {Strassmeier}~K.~G., {Collier-Cameron}~A., 1998, in Donahue~R.,
  Bookbinder~J., eds, {Tenth Cambridge Workshop on Cool Stars, Stellar Systems,
  and the Sun}.
\newblock {ASP Conference Series: 154}, San Francisco, p.~2073

\bibitem[\protect\citefmt{{Washuettl}, {Strassmeier} \&
  {Collier-Cameron}}{2001}]{washuettl2001cs11}
{Washuettl}~A., {Strassmeier}~K.~G., {Collier-Cameron}~A., 2001, in {Garcia
  Lopez}~R.~J., {Rebolo}~R., {Zapaterio Osorio}~M.~R., eds, ASP Conf. Ser. 223:
  11th Cambridge Workshop on Cool Stars, Stellar Systems and the Sun.
\newblock p.~1308

\end{thebibliography}

\end{document}